\documentstyle[12pt,aaspp4]{article}
\lefthead{Blum et al.}
\righthead{Galactic Center}

\tighten

\begin{document}

% ---------------defos-----------------------------------------------------

\def\zz{\hang\noindent}

\def\kms{km s$^{-1}$}
\def\pix{pix$^{-1}$}

\def\deg{$^{\circ}$}
\def\mic{{$\mu$m}}

\def\h2o{H$_2$O}

\def\ak{{\it $A_K$}}

\def\aple{$\mathrel{\hbox{\rlap{\hbox{\lower4pt\hbox{$\sim$}}}\hbox{$<$}}}$}
\def\apge{$\mathrel{\hbox{\rlap{\hbox{\lower4pt\hbox{$\sim$}}}\hbox{$>$}}}$}

% -------------------------------------------------------------------------
%\vglue 1.5in

\title{
%A Whole Bunch of Hot and Cool Stars in the
%Central 2$'$ of the Milky Way:
$JHKL$ Photometry and the $K-$band
Luminosity Function at the Galactic Center}

\author{R. D. Blum\altaffilmark{1,2,3}, K. Sellgren\altaffilmark{1,4}, and
D. L. DePoy\altaffilmark{1}}

\affil{Department of Astronomy, The Ohio State University\\
 174 W. 18th Ave., Columbus, Oh, 43210}

\altaffiltext{1}{Visiting Astronomer, Cerro Tololo Inter--American Observatory,
National Optical Astronomy Observatories, which are operated by the Association
of Universities for Research in Astronomy, Inc., under cooperative
agreement with the National Science Foundation.}

\altaffiltext{2}{Hubble Fellow}

\altaffiltext{3}{Current address: University of Colorado, JILA, Campus
Box 440, Boulder CO, 80309}

\altaffiltext{4}{Alfred P. Sloan Research Fellow}

%\vskip 100 pt
\begin{abstract}
$J$, $H$, $K$, and $L$ photometry for the stars
in the central $\sim$ 2$'$ ($\sim$ 5 pc) of the Galaxy are presented.
Using the observed $J-H$, $H-K$, and $K-L$ colors and
assumed intrinsic colors, we determine the interstellar
extinction at 2.2 \mic \ ($A_K$) for approximately 1100 individual stars.
The mean $A_K$ ($=$ 3.3 mag) is similar to previous results, but
we find that the reddening is highly variable and some stars
are likely to be seen through $A_K$ $>$ 6 mag.
The de--reddened $K-$band
luminosity function points to a significantly brighter component to
the stellar
population ($>$ 1.5 mag at $K$) than found in the stellar population
in Baade's window, confirming
previous work done at lower spatial resolution.
The {\it observed} flux of all Galactic center
stars with estimated $K_\circ$ (de--reddened magnitude) $\leq$ 7.0 mag
is $\sim$ 25 $\%$ of the total in the 2$'$ $\times$ 2$'$ field.

Our observations confirm the recent finding that several bright M stars
in the Galactic center are variable.
Our photometry also establishes the near--infrared
variability of the M1--2 supergiant, IRS 7.
\end{abstract}


\centerline{\it accepted for publication in the ApJ}

\section{INTRODUCTION}
The Galactic center (GC) stellar cluster is unique in the Milky Way.
The central core of stars in the Galaxy is an extremely
dense composite of older red giants and young, massive stars that exhibit
energetic winds through emission--line spectra.
Since the earliest work on the GC, investigators have been trying
to isolate these young and old(er) components.
In their discovery paper of the GC
stellar cluster, Becklin \& Neugebauer (1968) discussed a dominant
extended
infrared source and an infrared point source.
Infrared maps at increasingly higher angular resolution
resolved
the extended infrared source into a cluster of stars
(Becklin \& Neugebauer 1975; Allen, Hyland, \& Jones 1983;
Storey \& Allen 1983)
and identified the infrared point source of Becklin \& Neugebauer
(1968) as the brightest star in this cluster.
Infrared
spectroscopy of individual GC sources
(Treffers et al. 1976; Neugebauer et al. 1976;
Wollman, Smith, \& Larson 1982; Hall, Kleinmann \& Scoville 1982;
Lebofsky, Rieke, \& Tokunaga 1982, hereafter LRT) found that they
fell into two broad groups, sources with and without CO absorption
at 2.3 $\mu$m, a signature of late-type giants and supergiants.
Lacy, Townes, \& Hollenbach
(1982) suggested that a group of O and B stars could produce the
observed excitation in the diffuse ionized gas in one of the first
papers calling for a massive, young population to exist at the GC.
LRT identified
four stars as late--type
supergiants including the brightest 2 \mic \
source, IRS 7.
Using these as tracers of a star formation episode,
they postulated that a starburst (2 $\times 10^3$ M$_{\odot}$
within the central $\sim$ 5 pc) occurred approximately 10$^7$ yr ago.
Subsequent discoveries of one to two dozen
emission--line stars (Allen et al. 1990; Krabbe et al. 1991),
have increased the estimates of the starburst intensity
to 1.6 $\times 10^4$ M$_{\odot}$ (Krabbe et al. 1995).

Clearly, the GC represents a unique region of star formation in the
Galaxy. We need to account for the delivery and
subsequent collapse of large amounts
of gas in the central region. The massive stars currently identified
in the GC have lifetimes too short to have moved far from their birth
places (LRT; Morris 1993). We must also understand
the role of collisional processes that may result in unusual star
formation (Morris 1993) in the extremely dense GC region (\apge 10$^7$
M$_{\odot}$ pc$^{-3}$, Bailey 1980; Eckart et al. 1993; 1995), especially since
stellar collisions are likely to occur (Phinney 1989).

Star formation in the GC, despite its
unique history, can be studied by the same techniques used in other star
formation regions: determination of the distribution of stellar masses,
ages, and compositions.
As a first step, we present $J$, $H$, $K$, and $L$
photometry of the
stellar population in the central $\sim$ 2$'$ ($\sim$ 5 pc for a GC
distance of 8 kpc; Reid 1993). These data are used to investigate the observed
color--magnitude diagram and the de--reddened $K-$band luminosity
function.
Our $K-$band
luminosity function confirms the excess of bright
stars relative to the well studied bulge field, Baade's window (BW; $l$,$b$ $=$
0\deg,$-$4\deg),
pointed out previously by Lebofsky \& Rieke (1987), Rieke (1987, 1993),
Haller \& Rieke (1989) and Haller (1992).
The present data are of higher spatial resolution and reach fainter
magnitudes than these previous studies.

Of these excess bright stars, the blue
emission--line stars (Forrest et al. 1987; Allen, Hyland, and Hillier
1990; Krabbe et al. 1991; Libonate et al. 1995; Blum, Sellgren, \& DePoy
1995$a$; Blum, DePoy, \& Sellgren 1995$b$; Krabbe et al. 1995; Figer
(1995); Tamblyn et al. 1996)
have been associated with an epoch of star
formation \aple 10$^7$ yr ago, while the brighter, cool stars (LRT;
Sellgren et al. 1987; Krabbe et al. 1995)
may be associated with either the most recent epoch or with somewhat
older ones ($\sim$ 10$^8$ yr) as pointed out by
Haller (1992) and Krabbe et al. (1995).
A companion paper to this one (Blum, Sellgren, \& DePoy 1996; hereafter, Paper
II) presents
$K-$band spectra for some of the brightest cool stars in the GC.
The near--infrared photometry and
interstellar extinction estimates presented here are important to
the analysis of the 2.2 \mic \ spectra presented in Paper II. This combination
of photometry and 2.2 \mic \ spectra allows us to begin to delineate
the stellar components resulting from these different putative star formation
epochs.

\section{OBSERVATIONS AND DATA REDUCTION}
The primary data set for
the GC observations was obtained on the nights of 11 $-$ 13 July 1993 on the
4--m telescope at the Cerro Tololo Inter--American Observatory (CTIO)
using the Ohio State Infrared Imager and Spectrometer
(OSIRIS). OSIRIS is described
by DePoy et al. (1993).
All basic data reduction procedures were accomplished using
IRAF\setcounter{footnote}{4}.\footnote{IRAF is distributed by the National
Optical Astronomy Observatories.}

$J$ ($\lambda$ $\approx$ 1.25 \mic, $\Delta$$\lambda$ $\approx$ 0.24 \mic), $H$
($\lambda$ $\approx$ 1.65
\mic, $\Delta$$\lambda$ $\approx$ 0.30 \mic),
and $K$ ($\lambda$ $\approx$ 2.2 \mic,
$\Delta$$\lambda$ $\approx$ 0.40 \mic) images were obtained
for the $\sim$ central 2$'$
of the Galaxy using OSIRIS at $\sim$ 0.4$''$ pix$^{-1}$ during
the night of 13 July 1993.
The total exposure times were
320 s, 70 s, and 50 s at $J$, $H$,  and $K$, respectively.
Due to a small region of hot pixels on the OSIRIS NICMOS III
detector, the GC was offset 21$''$ W of center
to place it on a clean region of the
256$\times 256$ chip. The GC frames
were taken as a series with telescope offsets of 5$'' - 10''$ made
in between exposures to account for individual bad pixels.
Each of the eight $J$, seven $H$
and five $K$ frames was sky subtracted
with an image taken 500$'' - 600''$ off the GC and ratioed
by a dome flat--field image.
The seeing was approximately 1.0$''$ FWHM.
We find a plate scale of 0.390$''$ \pix  \ $\pm$ 0.007$''$ pix \
by adopting the positions
tabulated in Krabbe et al. (1995) for a number of bright, well
separated GC stars (IRS 1W, 6E, 9, 13, 16NE).

Secondary data used to investigate possible variable
stars and derive $K$ magnitudes for a small number of saturated
stars on the primary $K$ images
were obtained on 11 May 1993, 13 July 1993, and 24 April 1995.
In May 1993 we obtained $H$ and $K$ images of the GC with OSIRIS on the
the Perkins Telescope at Lowell Observatory near Flagstaff, Arizona.
The $K$ image was obtained with a $\sim$ 1 $\%$ neutral density filter.
The plate$-$scale was $\sim$ 1.6$''$ pix$^{-1}$.
We obtained narrow--band images of the GC on 13 July 1993
with OSIRIS on the 4m at CTIO (same plate$-$scale
as the primary data). Two $\Delta\lambda/\lambda$ $\sim$ 1 $\%$ filters
($\lambda \approx$ 2.19 \mic, $\lambda \approx$ 2.27 \mic) were
employed. A set of $J$ images was kindly obtained for us on 24 April 1995
by J. A. Frogel, again using OSIRIS at Lowell Observatory. These images were
obtained at $\sim$ 0.6 $''$ pix$^{-1}$ plate$-$scale.

\subsection{Photometry}
\subsubsection{Analysis}

The photometry in our primary data set was flux calibrated using seven stars
of known brightness on the CTIO/CIT photometric system
from a single field in BW (Frogel \& Whitford 1987, hereafter FW87).
These bright stars (B143, B145, B158, B159, B162, B163, and
B169, as denoted on the list in FW87)
were analyzed with aperture photometry,
and an aperture correction relating the instrumental and published
magnitudes for the flux standards was computed for each filter.
A second aperture correction relating the flux in this aperture to a
GC frame instrumental magnitude (instrumental magnitudes for the GC
frames were obtained with DAOPHOT; see below)
was calculated using between five and eight stars on the GC frames.
All aperture photometry used 5 pixel radius circular apertures.
For the BW standards, the seeing was approximately 1$''$ to 1.5$''$.
The 5 pixel aperture was large enough that corrections
due to seeing differences were insignificant.
The stars used in computing the second aperture correction on the
GC frames were chosen to be relatively bright and uncrowded.
The sum of these
two corrections was then applied to the GC photometry. The
photometric uncertainty (error in the mean)
due to the two corrections is $\pm$ 0.028,
0.030, and 0.024 mag at $J$, $H$, and $K$, respectively.

The narrow--band data in our secondary data set (used to obtain
photometry
for saturated stars in the primary $K$ data set, see below)
were flux calibrated using the derived magnitudes
for seven bright GC stars from Table 1 (IRS 6E, 11, 14NE, 15NE, 16NE, 28,
and OSU C2). These stars are represented roughly equally by hot and cool
stars (Table 1).
We find OSIRIS $K$ and OSIRIS narrow$-$band agree to $\pm$ 0.04 mag
(standard deviation in the mean, average of the two narrow$-$band
filters).
The secondary $H$ and
$K$ images (used to investigate variability only, see below)
were calibrated using the magnitudes for stars IRS 19, IRS 22,
star 74, and star 124 (Table 1); these four stars
agree to $\pm$ 0.06 and $\pm$ 0.08 mag  at $H$ and $K$, respectively
(standard deviation in the mean).
The secondary  $J$ images from April 1995
(also used to investigate variability) were calibrated using the same BW stars
as given above for the primary data set (uncertainty in calibration:
$\pm$ 0.04 mag.)
For these secondary $J$ images, magnitudes for IRS 7
were obtained by 5 pixel radius aperture photometry.

The primary $J$, $H$, and $K$ images and secondary narrow--band, $H$,
and $K$ images were analyzed individually with
DAOPHOT (Stetson 1987) in order
to obtain the stellar photometry. The particular
version of DAOPHOT employed is one modified by Jon Holtzman of Lowell
Observatory and integrated into the Ohio State/Lowell Observatory
VISTA image reduction and analysis
software program.
There are two main differences
between the version of DAOPHOT, as employed here,
and the original.
First is the use of a new grouping routine which is more
computationally efficient for extremely crowded fields. In this
routine, each star is analyzed separately with all of its neighbors which
lie within a critical radius. For the GC frames, a critical
radius of 10$-$12 pixels was used resulting in groups with
approximately up to 20 stars.
Second,
the background was determined locally and was a free
parameter in the profile fitting.
Each set of frames at a given wavelength was
analyzed with a point spread function defined by the same two or three stars.

The resulting DAOPHOT instrumental magnitudes for the primary data set
were then combined to
form average $J$, $H$, and $K$ lists.
The photometry lists from different frames were
merged by matching the coordinates of stars on the lists. After
making initial estimates of frame--to--frame offsets from bright stars,
the coordinate lists
were matched in an iterative procedure which typically resulted in
residual offsets between frames of 0.00 $\pm$ 0.20 pix.
Magnitudes from the individual lists were averaged and only
stars detected on two or more frames were kept in the final $J, H$, or
$K$ lists.
Figure~\ref{error} shows the DAOPHOT error as function of $J, H$, and $K$
magnitude. Brighter stars have DAOPHOT errors which are
comparable to the standard
deviation in the mean of measurements between different frames. Typical
DAOPHOT errors are 0.02, 0.03,
and 0.04 magnitudes for $J$ $<$ 15, $H$ $<$ 12, and $K$ $<$ 11 respectively.
As Figure~\ref{error} suggests, crowding results in larger DAOPHOT
errors than the standard deviation in the mean between frames for fainter
stars.
Average DAOPHOT errors are 0.18, 0.15, and 0.11 magnitudes
for all stars with $J$ $>$ 18, $H$ $>$ 16, and $K$ $>$ 13 respectively.
The $J, H$, and $K$ photometry lists were merged in the
same way as the individual frame lists with residual spatial offsets between
frames of 0.00 $\pm$ 0.30 pixels.

Nine stars with $K$ $<$ 8.6 mag appear saturated on the primary OSIRIS frames.
These saturated stars are represented with photometry
from the secondary data set narrow--band images (as indicated in Table 1).
However, the saturated stars are all tied to the primary OSIRIS flux scale
through the narrow--band images. We will argue later
that the GC {\it de--reddened} luminosity function has a brighter
component than that in BW. The brightest stars of this component rely
on the $K$ magnitudes derived from the narrow--band images.
The uncertainty associated with deriving $K$ magnitudes from
the narrow--band images could result in systematic differences
in magnitudes compared to the OSIRIS primary images.
These possible differences will not affect the
conclusions drawn later in this paper regarding the
$K-$band luminosity function in the GC because we know the brightest
stars were saturated and hence must be generally brighter than the stars in BW.
This is most easily seen by comparing the {\it observed} $K$ for the GC stars
to the BW stars with a typical value of extinction to the GC
applied (see Figure~\ref{obsklf} and section 2.3). Furthermore, the
number of stars not on the primary OSIRIS system is small relative to
the total number of stars in the bright component of the GC luminosity
function ($>$ 110 in either the observed or de-reddened luminosity
function; see section 3.4).

\subsubsection{Comparison to Other Data}
Color transformations are not available for
OSIRIS to the CTIO/CIT system for stars which are as red as those
in the GC ($J-K$ up to 7 mag).
However, our primary OSIRIS data can be compared to the CTIO/CIT GC photometry
derived from the images
of DePoy \& Sharp (1991, hereafter DS91).
DePoy \& Sharp originally presented a subset of their
photometry for a number of bright stars.
We analyzed their images (the coadded, not enhanced, images)
in the same way as for the OSIRIS data
described above.
IRS 7 was used as the PSF star for all
the DS91 images. For the $J$ and $H$ frames IRS 7 was nearer the
edge of the frame resulting in a PSF with a smaller radius. This
could affect the photometry at $J$ and $H$ for the
DS91 images; see the  appendix.

Figure~\ref{transJ} shows a comparison of the OSIRIS and DS91
data for 15, 16, and 15 stars at $J$, $H$, and $K$, respectively.
Only stars measured at $J$, $H$, or $K$ and with $J$ and $K$ mags
from DS91 as well were used. Suspected variables (see below)
and stars near the edge of the DS91 frames were excluded.
Each panel of this figure shows a weighted, least--squares fit as well.
The slopes for $\Delta J, \Delta H$, and $\Delta K$ vs. ($J-K$)$_{CTIO/CIT}$
are 0.003 $\pm$ 0.011, $-$0.023 $\pm$ 0.012, and $-$0.016 $\pm$ 0.009,
respectively. Figure~\ref{transJ} suggests no statistically significant
color transformation between CTIO/CIT and OSIRIS.
However, DS91 flux
calibrated their data using IRS 7 which we have now found to
be variable (see below).
For all stars matched at any of $J$, $H$, or $K$,
we find mean differences $\Delta J$, $\Delta H$, and $\Delta K$
of 0.06 $\pm$ 0.21, $-$0.09 $\pm$ 0.23, and $-$0.05 $\pm$ 0.18 mag,
respectively, when we compare Galactic Center
stars between DS91 and OSIRIS and exclude suspected variable stars.
The good agreement between the two sets of Galactic Center photometry
suggests that the magnitudes assumed by DS91 for IRS 7 (from Becklin et
al. 1978) were correct at the times of their observations, although we
cannot rule out the possibility that there is in fact a color term between the
OSIRIS and DS91 photometry, and that IRS 7 varied in such a way as to
mask this color term.
A photometry list of bright GC sources ($K$ $\leq$ 10.5), stars
with IRS numbers, and/or sources for which we have
obtained 2 \mic \ spectra (Paper II) is given in Table 1.
We have included
data from the DS91 images including their $L-$band (
$\lambda$ $\approx$ 3.45 \mic) measurements.

To summarize, we have derived $J$, $H$, and $K$ magnitudes for
stars in the central 2$'$ of the Galaxy.
The overall comparison to DS91, for which we have the
most stars in common of any data set at $K$, is good (rms difference of
0.18 mag for $\sim$ 50 stars in the most crowded central $\sim$ 15$''$
region). The comparison of a smaller set
of stars with measured $J$, $H$, and $K$ from the DS91 images and $J$, $H$, and
$K$ from the OSIRIS images
suggests that no significant color transformation exists between
the CTIO/CIT system (DS91) and the OSIRIS system (Figure~\ref{transJ}).

We have also compared our photometry
to recent values in the literature (see the appendix for details).
The OSIRIS data are in agreement with the high angular
resolution lunar occultation measurements of Simon et al. (1990) and
Simons et al. (1990) and consistent with the single source PSF fitting
of Tollestrup et al. (1989). OSIRIS $K$ magnitudes are systematically
faint compared to the aperture photometry of Rieke et al. (1989) and
Tamura et al. (1996) and the high angular resolution deconvolved
images of Eckart et al. (1993, 1995) as reported by Krabbe et al.
(1995). With the exception of the Krabbe et al. (1995) data, these
results are consistent with the type of photometry used in each
case, i.e. our
crowded field photometry is in agreement with the very high resolution
lunar occultation results and is systematically fainter than the previous
aperture photometry. Our results are conservative in the sense that
they are generally {\it fainter} than previous photometry, and we will
use them to show that a component of {\it brighter} stars exists in
the GC relative to BW.

\subsection{Variability of IRS 7 and Other Stars}
Our photometry shows that IRS 7 has varied in brightness at
$J$, $H$, and $K$ by approximately 0.8, 0.5, and 0.3 mag, respectively. We also
confirm the variability of IRS 9 and 12N found at $K$ by Tamura et al. (1996)
by finding $\Delta J$ of approximately 1.0 and 1.7 mag for IRS 9 and
12N, respectively, compared to previous photometry.

Table 2 details the photometry of IRS 7 for our primary and secondary
images. These data show that IRS 7 was brighter in July 1993
at $J$ and $K$ than previously (Becklin et al. 1978).
Both of the primary $H$ and $K$ measurements were
saturated for IRS 7 on our frames.
However, analysis of the two narrow--band images (near 2.2 \mic)
and secondary $H$ and $K$ images shows that
IRS 7 was also brighter at these wavelengths
relative to Becklin et al. (1978).
The IRS 7 $J$ magnitude derived from the 24 April 1995 images taken at Lowell
Observatory was consistent with its former value (13.8, Table 1).
The photometry of DS91 taken in 1989 September
and 1990 April provides another data point for IRS 7. DS91 used IRS 7
to calibrate their images, and since our comparison of the OSIRIS
photometry
with DS91 (Figure~\ref{transJ})
is consistent with no color transformation relative
to the CTIO/CIT system, it seems likely that IRS 7 had the same near
infrared magnitudes at the time of the DS91 and Becklin et al. (1978)
observations.
Tamura et al. (1994, 1996) reported evidence that the $K$ mag of IRS 7
may have brightened by $\sim$ 0.15 mag from 1991 to 1992, but found no
brightening when comparing data from 1991 and 1993. We have not added
the Tamura et al. data to Table 2 because it is tied only to aperture
photometry relative to IRS 1W, with no local sky subtraction.

IRS 9 and IRS 12N were
brighter at $J$ compared to DS91 (Table 2). These stars were
found to be
variable by Tamura et al. (1994, 1996),
so they, like IRS 7, were not included in the
color correction analysis discussed above. Tamura et al. (1996)
report a steady brightening of both IRS 9 and 12N from July 1991 to
August 1993.
We find the $H$ and $K$ magnitudes for these two stars to agree between
DS91 (1989 September) and OSIRIS (1993 July), but find that $J$
brightens
between DS91 (1990 April) and OSIRIS (1993 July), as given in Tables
1 and 2. It appears that IRS 9 and IRS 12N both became fainter sometime
between 1989 September and 1990 April, followed by a steady increase
in brightness until they had returned to their 1989 September
brightnesses
by 1993 July.

Table 1 shows both DS91 and OSIRIS photometry.
Twelve stars with $K$
measured from both data sets have $\Delta K$ $>$ 0.2.
Of these, four stars have
$\Delta K$ $>$ 3$\sigma$, and thus, may
have varied between the time in which the DS91 and OSIRIS data
were taken (IRS 1NE, 6E, 21, and star 87 $=$ IRS A19 for Tamura et al.
1996).
Note also that $H-K$ and $J-L$
are the only DS91 colors which are unaffected by possible variability.
Candidate variable stars from Tamura et al. (1996) and Haller (1992) are
identified in Table 1.

\subsection{Artificial Star Experiments}
Tests were conducted using artificial stars to assess our
completeness limits and effects of crowding on the $H$ and $K$ frames.
Because the GC is extremely crowded, complete artificial frames were
constructed rather than adding stars to the original data. For the
purpose of determining an estimate of the completion limit at $K$,
a luminosity function was constructed as the
sum of two components: the
observed
luminosity function for the bright stars and a renormalized BW
luminosity function (Tiede et al. 1995) for fainter stars (9 $\leq$ $K_{\circ}$
$\leq$ \ 15.0) with a mean reddening of $A_K = 3.5$ mag (see below) added.
Hereafter, magnitudes and colors with a ``$\circ$'' subscript will refer
to intrinsic, or de--reddened values.
The observed luminosity function and the BW relation are shown in
Figure~\ref{obsklf}.
The BW component was added to the observed luminosity function
such that it joined smoothly with the
observed luminosity function at $K$ $=$ 11.5. This composite function
was then approximated by a smooth power--law distribution (Log(N) $=$
0.35 $\times$ $K$ + const.).
The power--law
luminosity function and an assumed radial surface density distribution
($\Sigma$ $\sim$ $R^{-0.8}$, core radius of 4$''$, Becklin \& Neugebauer 1968;
Eckart et al. 1993; Eckart et al. 1995) were sampled randomly to
distribute stars on the artificial frame. Execution of DAOPHOT
in an analogous manner to that of the original frames suggests
that the GC $K$ frames are complete to $K$ \aple 12.
The completeness
limit may be less than this since the bright end of the test luminosity
function was taken from the observed luminosity function.
The input (power--law) luminosity function and the recovered one are shown in
Figure~\ref{kfake}. A similar test was made at $H$ and suggests the
$H-$band
images are complete to \aple 14.25.

To assess the affect of image crowding on the bright
end of the observed luminosity function,
ten frames were constructed as above but using only the reddened
BW luminosity function.
We added long period variables (LPVs)
to the Tiede et al. (1995) relation according to the
numbers added by spectral type given in FW87 (10 stars).
The BW luminosity function was normalized to have the same observed
luminosity as one of the GC frames, $L_K$ $=$ 0.95$\times$10$^6$
$L_{\odot K}$ ($R_{\circ}$ $=$ 8 kpc and $A_K$ $=$ 3.5, see below).
This value of the observed $K-$band luminosity corresponds to
an observed flux that is within 10 $\%$ of that
reported by Becklin \& Neugebauer (1968) for their 1.8$'$ diameter beam
measurement, accounting for the slightly larger OSIRIS frame area.
The BW relation was then fit by a smooth combination of third order
splines to generate the actual input artificial luminosity functions.
This fitting process resulted in input luminosity functions
with 9.9 $\leq$ $K$ $\leq$ 18.4. A total of 80,300 stars were distributed
on each frame with an average of 227 stars per frame at $K$ $\leq$ 10.9.
For the 10 frames, 7.1 stars, on average, were extracted with magnitudes
brighter than any star in the input luminosity function.
On average, 0.5 stars were up to
0.75 mag brighter than the brightest star in the input luminosity
function per frame, one star was up to 0.5 mag brighter per frame, and
5.6 stars were $<$ 0.25 mag brighter.
It is clear that we should
expect a small number of stars in our real frames to have similar
overestimates of their brightnesses. However, the number of such stars
is small and suggests
that we have not overestimated the $K$ magnitudes
for a significant number of the
brightest GC stars due to
chance alignments of bright stars.

\section{DISCUSSION}
\subsection{The Color--Magnitude Diagram}
The GC color--magnitude diagrams (CMDs)
are shown in Figures~\ref{cmdjhk}.
These diagrams clearly show the effects of strong and
variable interstellar extinction. The paucity of stars to the lower
right in each panel of Figure~\ref{cmdjhk} graphically demonstrates
the sensitivity of the observations to increased reddening: it is
difficult to detect faint, red stars at progressively shorter
wavelengths.
Differential reddening results in much larger color differences than
the intrinsic color
differences between hot and cool stars in the observed Galactic center
CMD, so that these
populations cannot be separated purely by the observed photometry. This is
shown in Figure~\ref{cmdjhk}
where we have over--plotted
the CMD of the old stellar population from BW (FW87) with values
of $A_K =$ 2 mag and $A_K =$ 4 mag, respectively. Except for the bright M
supergiant, IRS 7, the majority
of the GC CMDs are consistent with the CMD of the BW population.
The differences between the GC and BW become more apparent only when the
de--reddened $K-$band luminosity function (see below) or spectra of individual
stars
are considered (Paper II).

Figure~\ref{cmdjhk} also shows the
positions of hot and cool stars with spectral identifications.
Most hot stars appear more to the
blue in $H-K$ and (particularly) $J-K$.
This suggests, as expected,
that part of the dispersion in the observed CMDs is due to the mixed young and
old
populations.
Note that several possible hot stars such as
IRS 1W and IRS 6E with extremely red continua (Rieke et al.
1989; Libonate et al. 1995;
Blum et al. 1995$b$; Krabbe et al. 1995) are among the
reddest stars in the CMD.
Identifications for individual sources from spectra are
given in Table 1.

\subsection{Extinction and the Color--Color Diagram}

The $J-H$  vs. $H-K$ color--color diagram, Figure~\ref{ccdjhk},
allows us to estimate the interstellar extinction to individual stars.
Plotted along with the approximately 450 stars in Figure~\ref{ccdjhk} is
the interstellar reddening line based upon the interstellar extinction
curve of Mathis (1990) for which $E(J-H)$/$E(H-K)$ $\sim$ 1.6.
The majority of stars in the GC field lie along
this relation at positions corresponding to substantial $A_K$, suggesting
they are stars of normal colors seen through varying amounts of
interstellar extinction. For
intrinsic $J-H$ and $H-K$ of 0.7 mag and 0.3 mag (corresponding to
late type M giants, FW87) the majority of stars in Figure~\ref{ccdjhk}
lie at 2 mag \aple $A_K$ \aple 4 mag.

The intrinsic $J-H$ and $H-K$ colors of normal stars span
a relatively small
range in magnitude (e.g., Frogel et al. 1978 and FW87). For large values of
extinction, as indicated by Figure~\ref{ccdjhk}, relatively accurate
estimates of $A_K$ can be made by assuming a single pair of intrinsic
colors. We have calculated $A_K$ for the stars in Figure~\ref{ccdjhk}
which have $H-K$ within $\pm$ 0.5 mag of the reddening line
(375 stars)
by adopting intrinsic $J-H$ and $H-K$ of 0.7 mag and 0.3 mag and
using the Mathis (1990) interstellar extinction law. Here, we
take an average of $A_K$ as determined from the two colors.
The mean $A_K$ for stars with extinction determined in this way is
2.82 $\pm$ 0.71 mag.
For stars detected only at $H$ and $K$
(approximately 700 stars), \ak \ was
determined by de--reddening the star to an assumed intrinsic BW giant $H-K$
from FW87.
The mean $A_K$ for stars with extinction determined in this way is
3.58 $\pm$ 0.79 mag.

For stars detected only at $K$ (approximately
800), other techniques must be used to estimate \ak.
Figure~\ref{jhkobsklf} shows the {\it observed} luminosity function at
$K$ separately for stars detected at $J$, $H$, and $K$; for stars
detected at $H$ and $K$; and for stars detected only at $K$. These three
histograms are progressively shifted to fainter $K$ as would be expected
for stars intrinsically fainter at the same $A_K$. However, there is
also large overlap between the histograms. This would be expected for
stars of the same apparent $K$ brightness which are seen through larger
$A_K$. This effect was already demonstrated by the different values of
mean $A_K$
determined for the stars with $J$, $H$, and $K$ ($A_K$ $=$ 2.8 mag)
versus those with $A_K$
derived only from $H$ and $K$ ($A_K$ $=$ 3.6 mag).
The effect is also immediately apparent in the
middle panels of Figure~\ref{cmdjhk}. If the brightest stars detected
only at $K$ ($solid$ histogram of Figure~\ref{jhkobsklf}) have typical
intrinsic colors, then a lower limit for $A_K$ can be estimated by assuming an
$H$ magnitude equal to the completeness limit (14.25).
Stars detected only at $K$ must have $A_K$ $>$ 3.6 mag
for $K$ $<$ 11.6 mag, to avoid detection at $H$.
In this case,
stars detected only at $K$ would have $A_K$ of 6.5 mag for $K$ $=$ 9.25 mag,
5.9 mag for $K$ $=$ 9.75 mag, 5.3 mag for $K$ $=$ 10.25 mag, 4.6 mag for $K$
$=$ 10.75 mag, and
4.0 mag for $K$ $=$ 11.25 mag.
For the stars detected
only at $K$, therefore, a lower limit of $A_K$ $=$ 3.6 mag or the value of
$A_K$ derived
by assuming an $H-K$ using the $H-$band limiting magnitude, whichever was
greater, was adopted.

For the stars in Table 1 with $L-$band magnitudes from the DS91 images,
$A_K$ was computed from the same reddening law and an assumed
intrinsic $K-L$ of 0.2 mag (Johnson 1966).
If $K-L$ was within $\pm$ 0.5 mag of the reddening line
for $H-K$ vs. $K-L$ and no excess was indicated by the $J-H$ and $H-K$
colors,
this value was averaged with the other determinations (8 stars).
The average $A_K$ determined from the $K-L$ color (25 stars, including
stars with only $K$ and $L$) is 4.0 $\pm$ 2.0 mag.
This value is affected strongly by the stars IRS 2L, 3, 10EL, and 21
which had only $K-L$ measured. All are extremely red and may have
much larger intrinsic colors (e.g. due to circumstellar dust
emission) and hence smaller interstellar reddening. If these four are excluded,
the average $A_K$ determined from $K-L$ would be 3.3 $\pm$ 1.0 mag.

These individually derived reddening values were
used in constructing the de--reddened $K-$band luminosity function for
the GC (see below); results for all stars with $K$ $\le$ 10.5 mag, stars
with IRS numbers,
and stars for which $K-$band spectra are available (Paper II) in
Table 1.  The mean value of $A_K$ for all stars with one or
more observed near--infrared colors is 3.3 $\pm$ 0.9 mag.

\subsection{Stars with Infrared Excesses}

The color-color diagram (Figure~\ref{ccdjhk})
is also useful in identifying stars with potential excess emission.
Stars falling to the right of the reddening line by more
than 0.5 mag in $H-K$ are candidates for objects with excess emission.
This difference in $H-K$ from the reddening line is approximately
three times the difference between the reddest mean M giant ($H-K$)$_\circ$
in FW87
and the value we adopted for use in de--reddening the photometry.
$A_K$ was determined from $J-H$ only for these stars.
IRS 1W is a good example of this class:
it stands well off the reddening line in Figure~\ref{ccdjhk}
with an apparent infrared ``excess'', and
has a 2 $\mu$m spectrum which is extremely red (Blum et al. 1995b).

Becklin (1995) and Krabbe et al. (1995) have suggested that GC sources
with red, featureless 2 $\mu$m spectra and colors suggesting
an infrared excess are possibly young stellar objects (YSOs),
each still embedded in
its dusty cocoon and/or having an accretion disk which provides a
significant infrared excess.
IRS 1W is one example of a candidate YSO; another is IRS 21,
which has a spectrum similar to IRS 1W (Krabbe et al. 1995).
IRS 21 is barely visible on our $J$ and $H$ images, but is too
faint to be confidently extracted, in part due to its proximity
to other bright stars.
We place lower limits of $J$ $>$ 17.6 mag and $H$ $>$ 14.7 mag on IRS 21,
based on one pixel radius aperture photometry relative to nearby
IRS 33 (which may have similar background).
This implies $H-K$ $>$ 4.3 mag for IRS 21.

We have summarized published observations of
some luminous and well-studied YSOs in Table 3,
in order to compare YSOs to the {\it observed}
magnitudes and colors of YSO candidates in the GC (Table 1).
This comparison requires correcting the YSO magnitudes to
a distance of 8 kpc,
adding the foreground extinction toward the GC
($A_K$ = 3.6 mag),
and also subtracting any foreground extinction from the molecular
cloud surrounding the YSO.
It is difficult to correct YSOs for foreground extinction
because the intrinsic colors of YSOs are uncertain and very
model-dependent (Shu, Adams, \& Lizano 1987).
We have therefore made
two extreme assumptions about the intrinsic colors of YSOs: that either
all of the observed $H-K$ color of the YSO is intrinsic to the source
(red YSO), or all of the observed $H-K$ color of the YSO is due to foreground
extinction (blue YSO).

The $K$ magnitudes and $H-K$ colors of IRS 1W and IRS 21, and
other GC sources with IR excesses, are well matched by luminous YSOs
if the $H-K$ colors
of these YSOs are partially intrinsic and partially due to foreground
reddening.
If YSOs are intrinsically red, with their observed $H-K$ color
equal to their intrinsic $H-K$ color, then they would be
too faint and in many cases too red to account for the red, luminous GC
stars. If YSOs are intrinsically blue, with
their observed $H-K$ color entirely due to
foreground reddening, then they can easily account for the
observed magnitudes of red, luminous GC sources
with featureless spectra such as IRS 1W and IRS 21,
but their $H-K$ color would be too blue.
The truth probably lies someplace in between these extremes:
some of the YSO color is intrinsic, and some due to reddening,
making it plausible that YSO colors and magnitudes can match those of
red, luminous sources in the GC.

IRS 6E, another red object, would have
$(H-K)_{\circ}$ $=$ 2.3 mag if a typical value of $A_K$ (3.6 mag) is assumed.
This star has been identified
as a late type WR star (WC9) by Krabbe et al. (1995). The weak
emission lines detected by Krabbe et al. for IRS 6E (relative
to WC9 stars in the field) are consistent with a large excess.
WC8$-$9 stars have been identified in the field
with large excesses due to circumstellar dust emission.
Such stars may have weak or no infrared emission lines
as a consequence (Cohen et al. 1991). $JHK$ photometry for
dusty WCL stars (Williams et al. 1987; Cohen et al. 1991) and
$V$ and $R$ photometry to estimate $A_V$ (and hence $A_K$) suggest
that these stars have $(J-H)_\circ$ $=$ 0.76 mag to 1.8 mag and $(H-K)_\circ$
$=$
0.90 mag to 1.60 mag. If IRS 6E has $(H-K)_\circ$ $=$ 1.6 mag, $A_K$ would be
4.2 mag using the
observed $H-K$. The photometry of Williams et al. (1987) also suggests
that WC9 stars may have $(H-K)_\circ$ as small as 0.3 mag; therefore, for
consistency, we
have not adopted a different color for IRS 6E in Table 4 than any other
star even though its photometry and spectrum suggest it may be
intrinsically more red.
Other WC9 stars identified in the GC (Blum et al.
1995; Krabbe et al. 1995) also show weaker emission--lines than
field WR stars. These have smaller (yet non--zero) $(H-K)_{\circ}$
(see Table 4), so it is not clear what
may cause the observed dilution in these cases (source crowding is a
possibility).

\subsection{The $K-$band Luminosity Function}
Using the results of DAOPHOT crowded field photometry and the extinction
estimates above, we have constructed the de--reddened $K-$band luminosity
function (KLF) for the inner $\sim$ 2$'$ of the Galaxy (Figure~\ref{klum}).
As discussed in section 2.3,
due to extreme crowding at our spatial resolution (FWHM \aple 1$''$), the
luminosity function is only complete to $K_{\circ}$ \aple 8.5 mag.
Only stars with \ak \ $>$ 2 mag were included in the KLF in an attempt
to eliminate foreground stars (40 stars of 1100 with measured $A_K$
had \ak \ $<$ 2 mag). OSIRIS data were used in the KLF for stars
whose DS91 $K$ magnitude differed by
more than 0.2 mag from the OSIRIS $K$ magnitude
(approximately 15 stars; see section 2.1).

In Figure~\ref{klum}, we compare the KLF to a renormalized KLF
for the old stellar population in BW (FW87; Tiede et al. 1995).
The BW relation has been renormalized by requiring that
it account for the observed $K$ luminosity in the GC. For our $K$ images
(adding up the total observed flux on an image)
this corresponds to $L_K \approx 2.0\times10^7 L_{\odot K}$ assuming a mean
$A_K$ of 3.3 mag (this is an average $A_K$ for all stars for which
we calculated $A_K$ from one or more observed near--infrared colors)
and $R_{\circ} = 8$ kpc. This includes
a few percent correction which sets the darkest region of the
sky subtracted images
to zero flux. An alternate normalization scheme
gives a similar result:
assuming all the dynamically inferred mass
in the GC (Genzel, Hollenbach, \& Townes 1994, but corrected for the
projected mass within our $\sim$ 2$'$ field and taking R$_{\circ}$ = 8
kpc) is in a BW like population with
M/L = 1.2 M$_{\odot}$/L$_{\odot K}$ (Genzel et al. 1994) suggests
$L_K \approx 2.3\times10^7 L_{\odot K}$.

The comparison of the GC KLF and the renormalized BW KLF (Figure~\ref{klum})
shows an excess of bright stars at $K_{\circ}$ \aple 7 mag
which are presumably due to more recent star formation epochs.
The renormalized
BW KLF has stars as bright as $K_\circ =$ 5.5 mag while the GC KLF extends to
$K_{\circ}$ $\approx$ 2.0 mag.
The BW KLF has about 30 stars brighter than
$K_{\circ}$ $=$ 7.0 mag; the GC KLF has 149 stars in this range.
The artificial star experiments described in section 2.3 suggest roughly
3 $\%$ of the brightest stars in simulated GC images
might have observed magnitudes which
are too bright by 0.25 to 0.75 mag due to chance
alignment or image crowding, but this can still not
produce as many bright stars as are observed in the GC.

Stars with extremely red observed colors in the GC may be intrinsically redder
than we have assumed. This means their derived color excess
is too large, their $A_K$ is overestimated, and their de--reddened flux
is too bright.
In BW, the average LPV
has $(H-K)_\circ$ $\sim$ 0.6 mag (FW87). Such stars would have $A_K$
overestimated in our analysis by 0.5 mag. However, such stars are also
expected to be rare. Approximately 6 $\%$ of the stars in the FW87
luminosity function with $K_\circ$ $\leq$ 7.5 mag are LPVs.
A number of the brightest stars in Figure~\ref{klum} which have large
values of $A_K$ (and $K_\circ$ $\leq$ 7) estimated from only one color
(several using the $H-$band limiting magnitude)
have very red near--infrared spectra which are nearly featureless
or show emission lines (IRS 3, 6E, 21, 29N).
We have argued above that such stars may have excess emission and
redder intrinsic colors. If this is the case, then the estimated $A_K$
is too large and $K_\circ$ too bright.
IRS 3 has a value
of $A_K$ $=$ 9.94 mag which results in $K_\circ$ $=$ 0.84. This would imply
a $K$ luminosity which is much too high (consequently, IRS 3 is not
plotted in Figure~\ref{klum}). Two other stars detected only at $K$ and
$L$ and having no published spectra (IRS 2L and 10EL) also have large
$A_K$ and bright $K_\circ$ ($<$ 4 mag). This small number of stars
with potentially redder intrinsic colors does not change our conclusion
that there is a brighter component to the KLF in the GC than seen in
BW.

The fraction of {\it observed} flux for stars
with $K_\circ$ $\leq$ 7 mag is $\sim$ 25$\%$ of the total.
If we consider the integrated {\it de--reddened} flux of the brightest stars in
the
KLF, we find that those with $K_\circ$ $\leq$ 7.0
mag contribute approximately 65 $\%$ of the
total (excluding IRS 2L, 3, 6E, 10EL, 21, and 29N).
This larger percentage likely results from our sensitivity limits.
As stars are observed through larger $A_K$, only the
most luminous ones
will be detected.
The intrinsically less luminous stars at large $A_K$ are not
detected and so contribute
little to the de--reddened total flux which depends on the
mean $A_K$ of stars detected at two or more wavelengths.

Figure~\ref{ccdjhk}
suggests that many stars in the GC
have $A_K$ up to 4.0 mag. Since these stars all have
two measured colors and many fall on or near the reddening line, we can
be assured that some stars, at least, have reddening this high. If we
limit all stars in the KLF to $A_K$ $\leq$ 4.0 mag, then the fraction
of de--reddened flux in stars with $K_\circ$ $\leq$ 7 mag is approximately 30
$\%$.
For BW, the fraction of flux due to stars with $K_\circ$ $\leq$ 7.0 mag is 6
$\%$.

Thus, for even the most conservative limits on reddening,
there is strong evidence for a brighter component
to the GC KLF than exists in the KLF of the older population
in BW.
These excess stars, at least, we expect are the result of
more recent star formation epochs (i.e. their brightness suggests
they are more massive, and hence younger than the BW stars).
However, the majority of stars in this excess component do not yet have
spectroscopic identifications, so it is difficult to attribute
the excess to either the most recent starburst episode or an
older one(s). The more massive emission--line stars (see the
references in section 1) are relatively inconspicuous by their
estimated $K_\circ$ (Table 4) or observed $K$ (Table 1).
Of the emission--line stars
which have been identified spectroscopically and attributed to the
most recent star formation in the GC (Krabbe et al.
1995) only eight of 149 have $K_\circ$ $\leq$ 7 mag. Of these, two
(IRS 6E and 29N) have only one color
observed and may have infrared excess emission (see above and section
3.3) and, thus, estimated $K_\circ$ which are too
bright. By contrast, 18 cool stars with $K$ $\leq$ 7.0 mag
are identified by their spectra. Eleven of these have two colors
measured and all but one have $H$ and $K$, so their $A_K$ values
should be reliable. At least one of these stars
(IRS 7) is an M supergiant (LRT; Sellgren et al. 1987). The
remaining stars may be bright giants on the AGB (LPVs); some may be
supergiants. A major focus of Paper II will be an attempt to discriminate
between the red supergiants and potential AGB stars because these
stars trace different epochs of star formation. Two of the brightest
cool stars (IRS 9 and 12N) may be LPVs based on their $K-$band
spectra (Paper II) and large amplitude photometric variability (section 2.2).

The excess of bright stars at $K$ in the GC relative to BW
has important implications for the mass distribution
in the GC. Clearly, the  M/L ratio must change between BW and the
GC. This change affects the relative distribution of mass near the
GC in compact and extended components inferred from kinematics.
If the stellar population is
dominated by low mass stars, as in BW, but the light is enhanced by a
relatively few bright stars, the M/L ratio will be smaller than
typically assumed,
and the amount of mass inferred to be in a compact
object would be even greater (McGinn et al. 1989; Sellgren et al. 1990;
Krabbe et al. 1995; Haller et al. 1996).
On the other hand, recent star formation episodes biased toward
high mass star formation near the GC may have
resulted in forming a more compact cluster of stars and stellar remnants
super--imposed on the extended old population (Allen 1994)
resulting in a M/L ratio which is larger than typically assumed.
Our GC KLF only demonstrates the minimum excess of very bright stars. It
cannot be used to separate the entire young population, so it is
difficult to quantify this latter possibility. A recent analysis
of available surface brightness and kinematic data
(Saha, Bicknell, \& McGregor 1995) suggests an
extended mass distribution
with M/L \apge 2 inside 0.2 pc could explain the GC kinematic data
without a black hole.

An excess of luminous stars in the Galactic center has been
observed previously.
Lebofsky \& Rieke (1987), Rieke (1987, 1993), Haller \& Rieke (1989),
and Haller (1992)
reported an excess of luminous stars in
the Galactic center, relative to the luminosity function
in BW. Our data are of higher spatial resolution, and so
perhaps less susceptible to crowding problems (but by
no means completely unaffected).
In addition, we have established an upper--limit to
the contribution of the old stellar population in the GC by assuming its
KLF is similar to that for BW and accounts for all the dynamically
inferred mass in the GC.

DePoy et al. (1993), in their study of the KLF in
BW, showed that observations of this stellar population
at lower spatial resolution, corresponding to higher stellar surface
densities and/or larger distances, led to blending of groups of stars
which would then be falsely identified as single, more luminous stars.
The tests described in section 2.3
show that this is not  a significant problem in the GC
at our spatial resolution. Our experiments showed no significant number
of spurious detections of bright stars created by chance groupings.
Thus, the excess of luminous stars in the
KLF is real and not an artifact of image crowding.

\section{SUMMARY}
We have presented near--infrared photometry for approximately
2000 stars in the central $\sim$ 5 pc of the Galaxy. The $J-H$ vs. $H-K$
color--color diagram and $K$ vs. $J-K$, $H-K$ CMDs demonstrate
the large and variable interstellar extinction toward the GC.
Combinations of $J-H$, $H-K$, and $K-L$ colors
were used to estimate the near--infrared extinction, $A_K$, for approximately
1100 stars. Analysis of the observed colors
shows that the majority of stars are likely
to have intrinsic colors similar to bulge or field giants seen
through 2 mag to 4 mag of extinction ($A_K$).
While the mean $A_K$ for stars with one or
more observed colors is approximately 3.3 mag, we show that there are
likely stars for which $A_K$ is much higher ($A_K$ $>$ 6 mag in some cases).

Some GC objects may have excess circumstellar emission.
Potential excesses
are seen in stars which are possibly pre--main--sequence
objects (IRS 1W, IRS 21) and also in post--main--sequence objects
(IRS 6E), analogous to possibly similar objects elsewhere in the Galaxy.
The potential pre--main--sequence objects are compared to well studied
young stellar objects; these objects may have similar near--infrared
luminosities and colors as the GC objects depending on their circumstellar
vs. interstellar reddening.

Our $J$ band photometry confirm the variability of several stars
noted by previous investigators (IRS 9 and 12N),
and our $J$, $H$, $K$,
and 2.2 \mic \ photometry clearly establish the variability
of the well known M supergiant, IRS 7.

Our de--reddened photometry
was used to construct a $K-$band luminosity function which
confirms the excess of bright stars in the GC relative to the
old stellar population in Baade's window pointed out in previous work.
Our KLF is constructed from higher spatial resolution observations than
earlier work, and we demonstrate that the excess cannot
be due to image crowding. Approximately 25 $\%$ of the observed
flux in the GC comes from stars which comprise a brighter
component to the GC stellar population
than found in the old stellar population in Baade's window.
The majority of stars in this component ($K_\circ$ $\leq$ 7.0 mag) with
spectral identifications are cool stars. By contrast, the massive,
emission--line stars are less conspicuous in this component. It remains
to be seen whether the majority of the
brightest cool stars trace the most recent
star formation in the GC ($<$ 10 Myr) or somewhat older
star formation (\apge 100 Myr).
Two of the brightest cool stars
(IRS 9 and 12N) have near--infrared spectra (Paper II)
and photometric variations suggestive of LPVs.

This work was supported by National Science Foundation grants AST
90--16112, AST 91--15236, and AST 92--18449.
Support for this work was also provided by NASA through grant number
HF 01067.01 -- 94A from the Space Telescope Science Institute, which is
operated by the Association of Universities for Research in Astronomy,
Inc., under NASA contract NAS5--26555.
We wish to thank J. Holtzman for providing us with his modified
DAOPHOT routines and useful discussions regarding crowded field
photometry.
Our work has also benefited from discussions on
crowded field photometry with G. Tiede and L. Kuchinski.
We are grateful to J. Frogel for observing IRS 7 for us in April 1995.
We kindly thank M. Werner for communication of results prior
to publication. We also thank an anonymous referee whose comments have
resulted in a clearer presentation.

\section{APPENDIX}
Here we present details of our comparison
to the images of DS91
and compare our OSIRIS $K-$band photometry to recent values
given in the literature.
Comparison of all DS91 and OSIRIS magnitudes matched at $J$, $H$, or $K$
results in rms differences of 0.21 mag, 0.23 mag, and
0.18  mag for $\Delta J, \Delta H$, and $\Delta K$, respectively.
Here 4 of 28, 5 of 32, and 4 of 57 stars with differences greater than
2 sigma were excluded at $J, H$, and $K$, respectively.
At $J$ and $H$, the rms is about twice an average DAOPHOT error for
these same stars. At $K$
the rms is similar to an average DAOPHOT error. The difference
at $J$ and $H$ may be larger due to the smaller PSF radius
than for the $K$ frames. This would affect fainter
stars more. Plots of $\Delta J$ vs. $J$ and
$\Delta H$ vs. $H$ suggest that this is the case.
Therefore, we include the DS91 photometry in our analysis in the
following way: we have averaged the $K$ data from the
DS91 data set with our OSIRIS data set for stars which
have $\Delta K$ less than 0.2 for purposes of deriving $A_K$ and computing
the $K-$band luminosity function.
However, we present the observed DS91 and OSIRIS data separately in Table 1.
The DS91 $J$ and $H$ data were
used for estimating $A_K$ with OSIRIS data for
stars which have $\Delta K$ less than 0.2
if no OSIRIS $J$ or $H$ magnitude was measured.

We have compared our derived $K$ magnitudes with
the average $K$ magnitudes of Simons, Hodapp, \&
Becklin (1990) and Simon et al. (1990), and find excellent agreement
for the four bright IRS 16 sources (C, NE, NW, SW). These $K$
magnitudes were derived from high ($<$ 0.05$''$) spatial resolution lunar
occultation measurements. The difference between the OSIRIS $K$ magnitudes
and the average of the Simons et al. (1990) and Simon et al. (1990) data
(as reported by Simons et al. 1990)
is 0.02 $\pm$ 0.13 mag, where the uncertainty given is the
standard deviation.

The OSIRIS $K$ magnitudes are consistent, within
the uncertainties, with those presented by Tollestrup, Capps, \& Becklin
(1989), with OSIRIS fainter by 0.32 $\pm$ 0.60 mag (IRS 1NE, 1SE,
16NE, 16NW, 16SW, 16C, 16SW-E $=$ MPE+1.6$-$6.8 compared).
Tollestrup et al.
used single source PSF fitting to derive point source magnitudes from
their \aple 1$-1.5''$ images.

Tamura et al. (1996) present $K$ magnitudes for 26 stars in common with
the OSIRIS data set. Their 0.9$''$ synthesized aperture photometry is
systematically brighter than the OSIRIS data by 0.40 $\pm$ 0.40 mag
(comparing OSIRIS to the Tamura et al. August, 1993 data). The
uncertainty is the standard deviation, as above. For the large number of stars
compared, the uncertainty in the mean is considerably smaller ($\pm$
0.08 mag).
This is expected since the Tamura et al. data did not include background
subtraction and the synthesized apertures can suffer from contamination
by other stars (Tamura et al. were primarily looking for relative variations).

The OSIRIS $K$ magnitudes are
systematically fainter than those derived by Rieke, Rieke, \& Paul
(1989). For six stars in common
(IRS 10E, 10W, 13E, 16NE, 16NW, 16SW), we find a difference of 0.85
$\pm$ 0.30 mag. The Rieke et al. photometry was derived from synthesized
apertures on their low spatial resolution images (\aple 1.5$''$), which
is consistent with the somewhat brighter magnitudes.

We also find that the OSIRIS $K$ magnitudes are
systematically faint compared to those reported by Krabbe et al. (1995)
for 11 bright sources in the central $\sim$ 10$''$ (IRS 6E,
13E, 15SW, 16C, 16NE, 16NW, 16SW, 29N, 33E, AF,
MPE+1.6$-$6.8). The difference in $K$ for these
11 sources is 0.80 $\pm$ 0.31 mag. Comparing only the four bright IRS 16
sources, as for the lunar occultation measurements, the difference
between OSIRIS and Krabbe et al. (1995) $K$ magnitudes is 0.83 $\pm$ 0.22 mag,
Krabbe et al. again being brighter.  Krabbe et al. (1995) actually
report $K$ magnitudes derived from the high angular resolution (deconvolved
resolution $\sim$ 0.2$''$) images of Eckart et al. (1993,
1995); it is not clear how the flux scale was calibrated or
whether the magnitudes are affected by the Eckart et al. image
restoration technique. If IRS 7 was
used as the flux calibrator, then it is possible that this could result
in some of the difference, as we find that IRS 7 is variable (see
below). Krabbe et al. (1995) do note that their absolute $K$ magnitudes
appear to be 1$-$2 magnitudes brighter than expected for stars of
similar spectral type elsewhere in the Galaxy.

\newpage

%FIGURES

\begin{figure}
%\plotfiddle{f1.eps}{6.5 in}{0}{70}{70}{-218}{-20}
\caption[]{DAOPHOT errors for $J$, $H$, and $K$ magnitudes.
The errors include the uncertainty associated with the aperture
corrections to the instrumental magnitudes.
For bright
stars, typical uncertainties reported from DAOPHOT were similar to the
observed scatter between frames. Fainter stars show the effects of
crowding as well as photometric uncertainty; see text.}
\label{error}
\end{figure}

\begin{figure}
%\plotfiddle{f2.eps}{6.5 in}{0}{70}{70}{-218}{-20}
\caption[]{Comparison of OSIRIS and CTIO/CIT photometry.
The dashed lines
are weighted fits to the data.
This comparison of the brightest and reddest
stars in common between the OSIRIS frames and the GC photometry
derived from the images of DS91 suggests no statistically significant
color correction is warranted (but see text for a discussion
on the flux calibration of DS91 using IRS 7).
The bluest stars are giants of known magnitude in BW.}
\label{transJ}
\end{figure}

\begin{figure}
%\plotone{f3.eps}
\caption[]{The observed $K$ Band (2.2\mic) luminosity function ({\it
solid histogram}) and BW luminosity function ({\it dotted histogram}).
The BW relation has been shifted by applying a mean extinction
of $A_K$ $=$ 3.5. The two luminosity functions were combined
at $K=11.5$ to create an artificial luminosity function for
use in estimating the completeness limit of the $K-$band data.
The BW component was normalized so that it joined smoothly
with the observed luminosity function.
See the discussion in the text and Figure~\ref{kfake}.}
\label{obsklf}
\end{figure}

\begin{figure}
%\plotone{f4.eps}
\caption[]{The artificial $K$ Band (2.2\mic) luminosity function ({\it
solid histogram})
created by combining the observed GC luminosity function
for $K$ $<$ 11.5 and
the BW LF (Tiede et al. 1995), reddened by $A_K$ = 3.5, for $K$ $>$
11.5. The two luminosity functions
in Figure~\ref{obsklf}
were combined and then fit by a power--law. The power--law was randomly
sampled to produce the artificial luminosity function shown here.
The recovered histogram ({\it solid triangles}) suggests the
GC $K-$band data is complete to $K$ \aple 12.}
\label{kfake}
\end{figure}

\begin{figure}
\caption[]{
Observed color--magnitude diagrams for the $\sim$ 2$'\times2'$ field of
the Galactic center. The three pairs of plots show the same
observed Galactic center photometry ({\it small filled circles}) with different
overlays.
The very red colors and large dispersion in $J-K$ and  $H-K$
(top two panels) demonstrate the strong and variable
interstellar extinction toward the Galactic center.
The importance of obtaining spectra is demonstrated by the fact that
based solely on observed magnitudes and colors, all but one of the
Galactic center stars in the CMD is consistent with an old stellar population
like that in Baade's window (middle two panels),
after interstellar extinction values of
$A_K$ $=$ 2 mag ({\it open circles}) and $A_K$ $=$ 4 mag ({\it open triangles})
are applied to the Baade's window data to
match the estimated extinction toward the Galactic center.
Data for Baade's window was
taken from Frogel \& Whitford (1987).
Galactic center stars with
spectral classifications (bottom two panels) from near--infrared spectra are
identified as either hot ({\it large filled triangles})
or cool ({\it large filled circles}) stars.}
\label{cmdjhk}
\end{figure}

\begin{figure}
%\plotone{f5.eps}
\end{figure}

\begin{figure}
%\plotfiddle{f6.eps}{6.5 in}{0}{70}{70}{-218}{-20}
\caption[]{$J-H$ vs. $H-K$ color-color diagram. The dashed line
represents the interstellar extinction law of Mathis (1990) for which
$E(J-H)$/$E(H-K)$ $\sim$ 1.6. Stars
which fall to the right of the reddening line by more than
0.5 mag in $H-K$
may have intrinsic excess; see
text.}
\label{ccdjhk}
\end{figure}

\begin{figure}
%\plotone{f7.eps}
\caption[]{
Observed $K-$band luminosity function for stars detected
at $J$, $H$, and $K$ ({\it dashed histogram}); for stars detected
only at $H$ and $K$ ({\it dotted histogram}); and for stars detected
only at $K$ ({\it solid histogram}). Stars detected at $J$, $H$, and
$K$ have mean $A_K$ (2.8 $\pm$ 0.7)
smaller than that of those detected only at $H$ and $K$ (3.6 $\pm$ 0.8).
Similarly, some stars detected only at $K$ are probably intrinsically
luminous stars which are more heavily reddened
($A_K$ up to 6.5) than
the mean value for those detected at $H-K$; see discussion in text.}
\label{jhkobsklf}
\end{figure}

\begin{figure}
%\plotone{f8.eps}
\caption[]{
De--reddened $K-$band luminosity function for the central
$\sim$2$'$ of the Galaxy ({\it solid histogram}).
The Galactic center
shows a significant excess of bright stars relative to the renormalized
Baade's window
({\it dashed line}) luminosity function.
The normalization of the Baade's window luminosity function
is based on the observed $K$ luminosity of the GC and is
also consistent with putting all the dynamically observed mass
of the GC into a population with a mass--to--light ratio
like that of the population in  Baade's window; see text.}
\label{klum}
\end{figure}

\end{document}